\documentclass[conference]{IEEEtran}
\IEEEoverridecommandlockouts
\usepackage{cite}
\usepackage{amsmath,amssymb,amsfonts}
\usepackage{algorithmic}
\usepackage{graphicx}
\usepackage{textcomp}
\usepackage{xcolor}
\usepackage{listings}
\usepackage{wrapfig}
\usepackage{multicol}
\usepackage{subcaption}
\usepackage{tikz}
\usepackage{amssymb}
\usepackage{comment}
\usepackage{enumitem}

\newcommand{\linenumber}[1]{#1:}

\definecolor{codegreen}{rgb}{0.0,0.45,0.1}
\definecolor{codegray}{rgb}{0.5,0.5,0.5}
\definecolor{codepurple}{rgb}{0.58,0,0.82}
\definecolor{codeblue}{rgb}{0,0,1}
\definecolor{codered}{rgb}{1,0,0}
\definecolor{codebrown}{rgb}{0.6,0.3,0.1}
\definecolor{backcolor}{rgb}{0.99,0.99,0.99}
\definecolor{codecyan}{rgb}{0,0.6,0.6}

\lstdefinelanguage{csl}{
    morekeywords={task, if, else, == },
    keywordstyle=\color{codebrown},
    classoffset=1,
    morekeywords={@fmovs, @fadds, @fmacs, @increment_dsd_offset, @set_dsd_base_addr},
    keywordstyle=\color{codepurple},
    classoffset=2,
    morekeywords={u16, u32, f32, void},
    keywordstyle=\color{codeblue},
    classoffset=0,
    comment=[l]{//},
    commentstyle=\color{codegreen}\ttfamily,
    stringstyle=\color{codered}\ttfamily,
    morestring=[b]',
    morestring=[b]",
    backgroundcolor=\color{backcolor},
    basicstyle=\footnotesize\ttfamily,
    tabsize=2
}

\def\BibTeX{{\rm B\kern-.05em{\sc i\kern-.025em b}\kern-.08em
    T\kern-.1667em\lower.7ex\hbox{E}\kern-.125emX}}
\begin{document}

\title{Exploring Sparse Matrix Multiplication Kernels on the Cerebras CS-3}

\author{\IEEEauthorblockN{Milan Shah}
\IEEEauthorblockA{\textit{Dept. of Electrical and Computer Eng.} \\
\textit{North Carolina State University}\\
Raleigh, NC \\
mkshah5@ncsu.edu}
\and
\IEEEauthorblockN{Sheng Di}
\IEEEauthorblockA{\textit{Computer Science Division} \\
\textit{Argonne National Laboratory}\\
Lemont, IL \\
sdi1@anl.gov}
\and
\IEEEauthorblockN{Michela Becchi}
\IEEEauthorblockA{\textit{Dept. of Electrical and Computer Eng.} \\
\textit{North Carolina State University}\\
Raleigh, NC \\
mbecchi@ncsu.edu}}

\maketitle

\pagestyle{plain}

\begin{abstract}
In recent years, novel AI accelerators have emerged as promising alternatives to GPUs for AI model training and inference. One such accelerator, the Cerebras CS-3, has demonstrated strong performance on machine learning as well as scientific applications, such as molecular dynamics and seismic simulations. While the benefits of Cerebras systems for dense workloads have been well demonstrated, their potential for sparse workloads is not yet understood, particularly for large matrices that cannot fit on the device when represented in dense format. Yet, many applications, such as linear solvers and graph neural networks, rely on large sparse matrices.

In this work, we make a step toward a better understanding of the use of CS-3 platforms for sparse operations. To this end, we explore two key sparse linear algebra kernels, sparse-dense matrix multiplication (SpMM) and sampled dense-dense matrix multiplication (SDDMM), on the Cerebras CS-3. We propose low-level CS-3 designs for these operations and optimize them to improve I/O performance, memory footprint, and scalability to large matrices. We evaluate speedup relative to the CPU. The results show that our CS-3 kernels can outperform the CPU by up to 100$\times$ for SpMM on 90\% sparse matrices, with performance improving as matrix dimensionality increases. SDDMM on the CS-3 can outperform the CPU by up to 20$\times$ on 90\% sparse matrices. However, as sparsity increases beyond 99\%, our kernels suffer performance degradation, approaching the performance of state-of-the-art CPU libraries or even underperforming them.
\end{abstract}


\section{Introduction}
Novel AI accelerators have emerged in recent years as tools for accelerating machine learning pipelines and have also demonstrated their potential beyond machine learning. For example, Cerebras systems have demonstrated good performance on a variety of scientific workloads, including molecular dynamics simulations~\cite{10793165}, seismic simulations~\cite{ltaief2023scaling}, Monte Carlo particle transport~\cite{montecarlo2024}, and stencil computations~\cite{stencil_sc24, 10.1145/3779212.3790124}.

These accelerators attempt to address the limitations of GPU architectures in the context of DNN/LLM training and inference. While GPUs can enable massive SIMD parallelism across many cores, communication between on-chip and off-chip memory between kernel calls can bottleneck DNN training. The architects of new AI accelerators, such as the Cerebras CS-3 \cite{cerebras_product}, SambaNova RDU \cite{sn30_product}, Graphcore IPU \cite{ipu_product}, and Intel Habana Gaudi \cite{intel_gaudi}, have opted to greatly increase on-chip memory capacity in an effort to reduce frequent exchanges with off-chip memory involving data structures such as activations and gradients during DNN training. Additionally, these accelerators exhibit a range of architectures: for example, the Graphcore IPU is composed of 1472 parallel compute units that can execute their own instruction stream (Multiple-Instruction-Multiple-Data architecture), while the CS-3 and RDU are dataflow architectures, where computation is spatially mapped to physical compute units and results from one compute unit are directly fed to the next unit via high-speed on-chip interconnects \cite{cerebras_product,sn30_product}.

In this work, we study the ability of  CS-3 systems to perform sparse-dense matrix multiplication (SpMM) and sampled dense-dense matrix multiplication (SDDMM), and explore optimizations to improve host-device data transfers for these computations on the CS-3. SpMM and SDDMM are popular sparse linear algebra kernels used in a wide variety of applications, such as pruned deep neural networks, graph neural networks, recommendation systems, scientific simulations, linear solvers, sparse attention in transformers, electronic design automation, and graph analytics \cite{sparsifieddnnmodels, spmmlinearsolvers, sddmmusecases, gcnpaper, gat_paper}. However, the current sparsity support offered through Cerebras high-level programming interfaces, such as CSTorch, is inadequate for large matrices, especially those that cannot fit on the device when represented in dense format.

Our key contributions are as follows:
\begin{itemize}[leftmargin=*]
\item We propose SpMM and SDDMM kernels written in the low-level Cerebras Software Language (CSL) for efficient execution on the CS-3. Our kernels operate on sparse matrices represented in sparse storage formats similar to CSR and SELLPACK.
\item We optimize our SpMM implementation to support multiple I/O channels for streaming data from the host to the CS-3 and to enable the efficient processing of large matrices.
\item We evaluate our implementations on a variety of synthetically generated sparse matrices to characterize CS-3 performance relative to the CPU, discussing both the performance potential of the CS-3 and the bottlenecks of this hardware.
\end{itemize}
Our evaluation suggests that, for large matrices, SpMM on the CS-3 can achieve up to 100$\times$ speedup over  CPU for matrices with 90\% sparsity. SDDMM performance is typically an order of magnitude faster than the CPU, scaling at the same rate as the CPU as density increases. As matrices become hyper-sparse, our CS-3 kernels suffer from high host-device communication overhead relative to computation, approaching baseline CPU performance.
Further avenues of exploration, such as different SpMM/SDDMM designs tailored to hyper-sparse matrices, different mapping and data distribution strategies, as well as data transfer and reuse policies suited to applications with back-to-back matrix operations, might enable more effective use of these devices for sparse workloads.

\section{Background}
\subsection{SpMM and SDDMM}
\textbf{SpMM}, or sparse-dense matrix multiplication, is a fundamental sparse linear algebra routine common in graph neural networks \cite{gcnpaper, gat_paper}, pruned deep neural networks \cite{sparsifieddnnmodels}, and linear solvers \cite{spmmlinearsolvers}. Formally, SpMM performs the following computation:
\vspace{-2mm}
\begin{equation}
    Y = AX, A\in \mathbb{R}^{K \times N},X\in \mathbb{R}^{N \times D}
    \vspace{-2mm}
\end{equation}
$A$ is the sparse matrix, $X$ is the dense matrix, and $Y$ is the dense output matrix. In many applications (such as GNNs), $K=N$, thus $A$ is square. $N$ can vary significantly from application to application, but often is on the scale of millions to billions for real-world graphs.

\textbf{SDDMM}, or sampled dense-dense matrix multiplication, is a kernel used in graph attention networks \cite{gat_paper}, recommendation systems, and scientific computing \cite{sddmmusecases}. Formally, SDDMM performs the following computation:
\vspace{-2mm}
\begin{equation}
    Y=A \odot(BC), A\in \mathbb{R}^{N \times N},B\in \mathbb{R}^{N \times D}, C\in \mathbb{R}^{D \times N}
    \vspace{-2mm}
\end{equation}
As is the case with SpMM, $N$ can be extremely large depending on the application, ranging from millions to billions. SDDMM essentially only computes output elements of $Y$ where there exists a nonzero in $A$.

\subsection{Cerebras CS-3}
\begin{figure}[t]
\centerline{\includegraphics[width=0.9\linewidth]{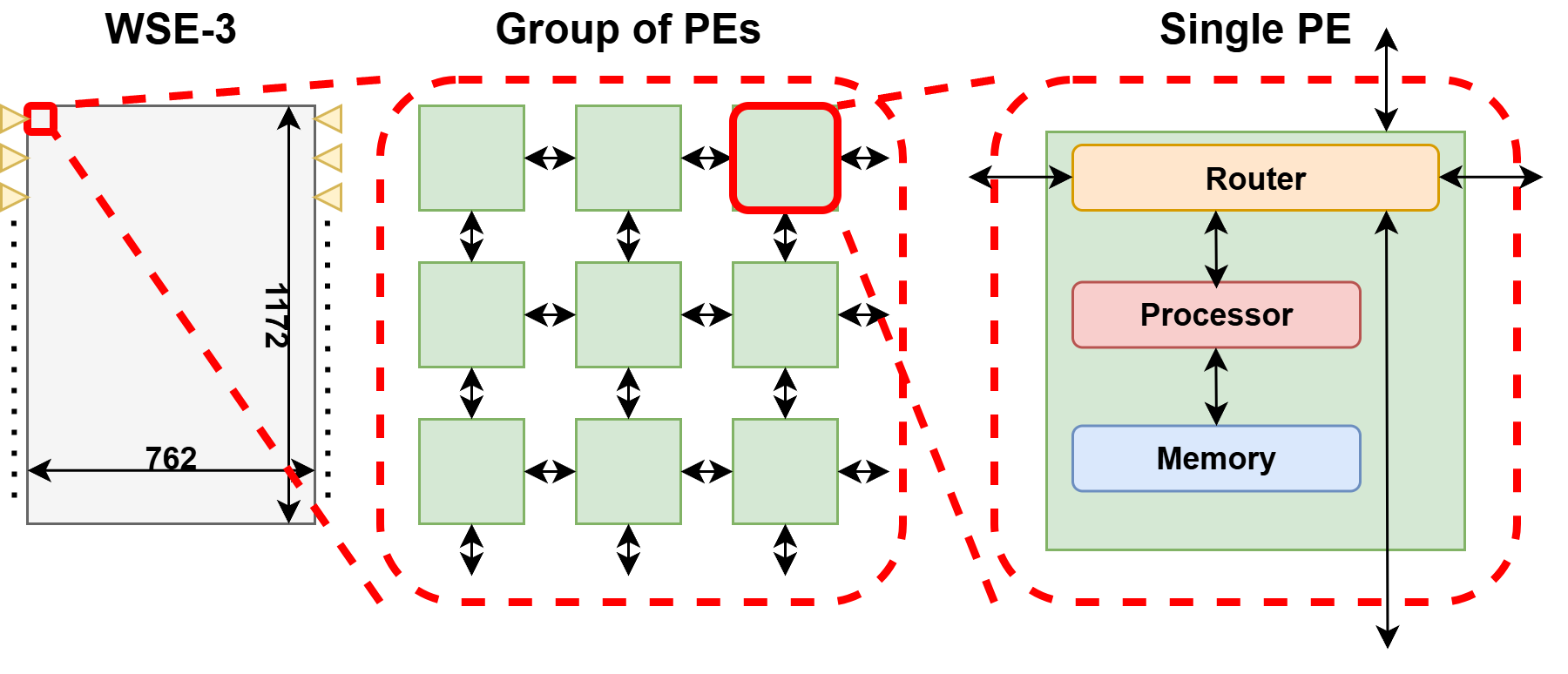}}
\caption{The Cerebras Wafer-Scale Engine 3 (WSE-3, or CS-3 for short) high-level architecture overview. The left-most rectangle represents the full chip, with I/O channels along the west and east edges (yellow triangles) and 762 $\times$ 1172 processing elements (PEs). The middle region of the diagram illustrates a group of 9 PEs and the right region shows the internal components of a single PE. PEs are connected to their cardinal direction neighbors.}
\label{fig:cerebras}
\vspace{-1mm}
\end{figure}

Cerebras's Wafer Scale Engine (WSE) architecture is summarized in Fig.~\ref{fig:cerebras}. The wafer-scale chip offers a grid of programming elements, or PEs—approximately 900,000 on CS-3 systems. Each PE consists of three components: a processor core, 48 KB of local memory, and a router that is directly connected via bidirectional links to its own processor and to the routers of the four nearest neighboring PEs. By providing access to the on-wafer interconnect, the router enables communication with other PEs. Once loaded in local memory, data can reside there for the duration of computation. 32-bit data, called wavelets, can move from one PE to a directly connected neighbor in 1 clock cycle. Overall, CS‑3 offers about 44 GB of on-chip memory and an aggregate on-chip memory bandwidth of approximately 21 PB/s. PEs are capable of SIMD instructions, and the SIMD width depends on the hardware instruction and WSE generation. WSE systems are typically paired with large external memory units called MemoryX, with typical installations ranging from 1 TB to a few terabytes in capacity.

CS can be programmed in two ways: using high-level PyTorch and TensorFlow interfaces, or using a low-level software development kit (SDK), which includes a host library that can be invoked from Python programs and the CSL language—a C-like language for writing custom software kernels for the WSE~\cite{csl-doc}. The host library allows transferring data between host and device memory and invoking kernels on the device (i.e., the WSE). The CSL language includes special Data Structure Descriptors (DSDs), which allow accessing contiguous memory locations (possibly with strided accesses) and executing SIMD instructions on them. In addition, a class of DSDs (called fabric DSDs) allows explicitly sending wavelets to other PEs along predefined routes. CSL also supports multithreading and allows instantiating different kinds of tasks, including ``data tasks'' that are activated upon reception of a wavelet. The language provides support for buffers (e.g., FIFOs) and asynchronous operations to improve efficiency. Programmers can configure the routes and the code executed by the PEs through a special ``layout file,'' which can be written using either CSL or an SDK API, providing fine-grained control over PE code assignment and fabric configuration. 

While Cerebras' PyTorch interface offers some support for sparsity~\cite{cerebras_sparsity}, it still requires models/matrices to be stored in a dense format, with an additional \textit{sparsity mask} indicating the location of zero values. This restriction limits the size of matrices that can be supported, typically to small weight matrices from DNNs and transformers, as opposed to large matrices used in graph neural networks and scientific applications. Thus, in this work, we design custom sparse kernels using the CSL low-level language.

\subsection{Distributed matrix multiplication approaches}
There are several approaches to distribute a matrix multiplication in distributed environments. The approaches depend on the placement of the source and result matrices and their distribution on the system. Traditionally, borrowing from the dense case, distributed matrix multiplication algorithms are categorized as 1D, 1.5D, 2D, 2.5D, and 3D decompositions, depending on the data distribution and, consequently, the communication volume they entail. At a high level, as the decomposition dimension increases, memory usage per PE may increase due to data replication, while the overall communication cost decreases because more computation can be performed locally. Without replicating matrix $X$, our SpMM implementation described in Section~\ref{sec:design:spmm} corresponds to a 1.5D decomposition, in which $A$ is conceptually replicated only along processor columns during streaming, while $X$ is not replicated. Streaming matrix A avoids on-chip storage of the entire matrix. While the ratio of computation to communication costs on CS-3 is higher than in traditional distributed systems, \textit{limiting communication} remains important for performance.

\subsection{Related Work}
While less studied than dense-dense or sparse-sparse kernels (such as GEMM and SpGEMM), SpMM has recently received increased consideration, in part due to its use in machine learning. To this end, several works have explored efficient implementations on GPUs~\cite{spmm-gpu1, spmm-gpu2, spmm-gpu3, spmm-gpu4, spmm-gpu5, spmm-gpu6}, in some cases in combination with SDDMM~\cite{spmm-sddmm-gpu1,spmm-sddmm-gpu2}. Overall, these papers propose techniques aimed at improving memory coalescing, decreasing warp divergence (through load balancing), improving data locality for better reuse, and leveraging tensor cores on GPU. Because the architectural features of GPUs and CS differ significantly, many of these optimizations do not generalize well to CS.

CS can be considered a \textit{distributed system on a chip}, with hundreds of thousands of cores connected via a high-bandwidth on-wafer network. Thus, existing research on distributed-memory sparse kernels, such as~\cite{spmm-distr1, spmm-distr2, spmm-distr3, spmm-distr4}, is particularly relevant to this work. These studies focus on reducing communication costs (through strategies such as different data distributions across nodes, efficient collective communication primitives, and overlapping communication with computation), load balancing, and improving memory efficiency via blocking, the use of sparse formats, and data streaming. While the communication costs and the cost of communication relative to computation are very different on CS-3, here we revisit some of these techniques (streaming, sparse matrix formats) on this novel accelerator.

A recent technical report~\cite{spmm_wse2_2025} explored the use of different sparse matrix formats for deploying SpMM on CS using the CSL language. However, that work targeted sparse matrices used in DNNs and LLMs, which are significantly smaller and less sparse than the adjacency matrices found in other applications, such as GNNs. Here, we explore large matrices and the effect of sparsity on performance. 

Recent work~\cite{song2022sextans, Liu2024FPGASpMM} has explored FPGA-specific optimizations for SpMM, including on-chip buffering, streaming, workload balancing, nonzero scheduling, high-bandwidth memory integration, and sparse data formats. While both FPGAs and Cerebras systems are spatial architectures with dataflow execution, they have fundamental differences: for example, FPGAs have limited on-chip memory, slower interconnects, and require low-level hardware design, whereas Cerebras systems provide large low-latency on-chip memory, a high-bandwidth, programmable interconnect, general-purpose PEs, and higher-level programmability, reshaping the optimization space.

\section{Design}

\subsection{SpMM on CS-3}
\label{sec:design:spmm}

We present our SpMM implementation in two steps. First, we discuss a basic SpMM design and its rationale and limitations (Section~\ref{sec:spmm:basic}). Then, we introduce two optimizations aimed at improving host-device communication and reducing serialization costs (Section~\ref{sec:spmm:optimized}).

\subsubsection{Basic SpMM Design}
\label{sec:spmm:basic}

\begin{figure}[t]
\centerline{\includegraphics[width=0.8\linewidth]{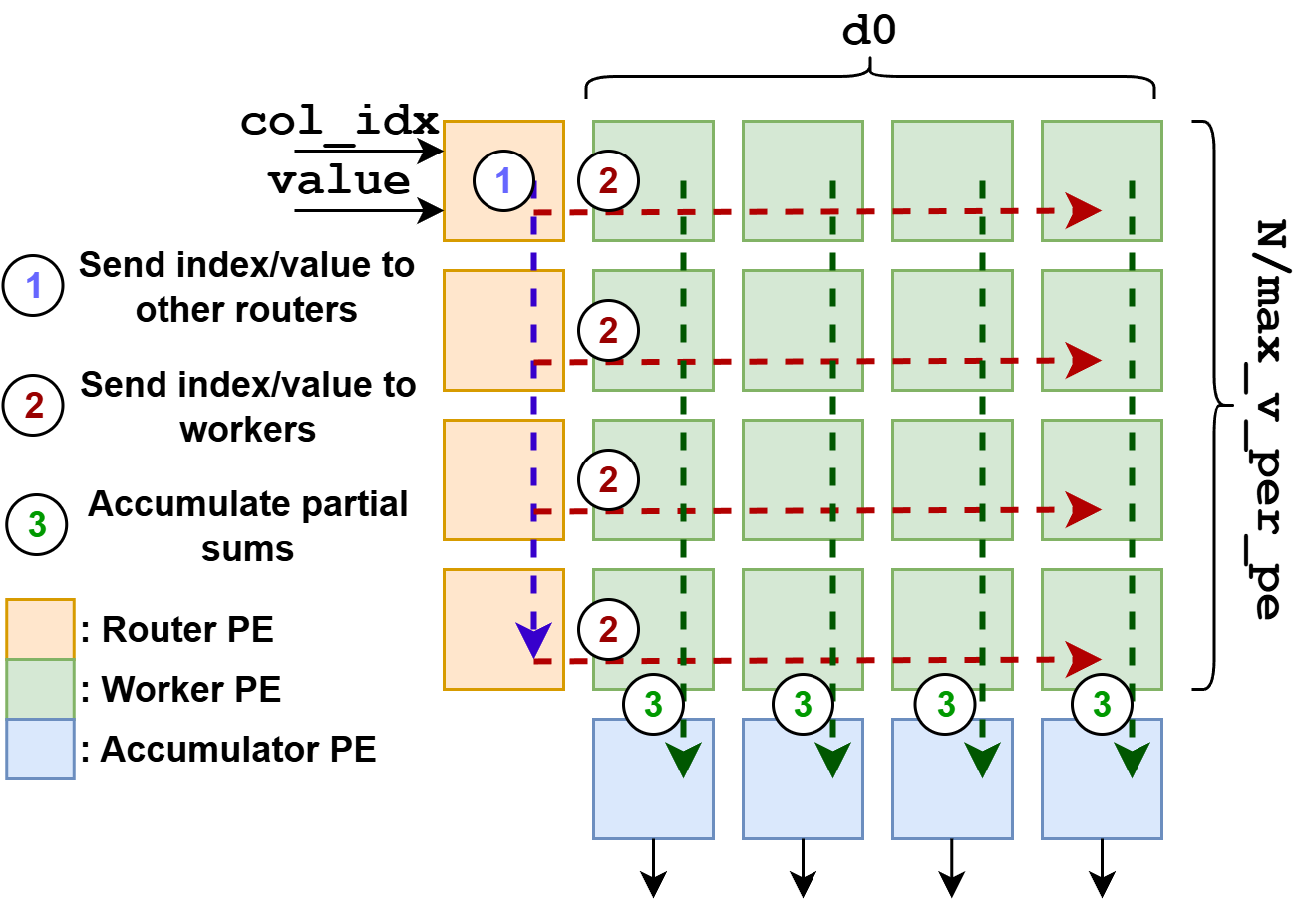}}
\caption{Basic SpMM Design ($D = d_0$)}
\label{fig:base_spmm}
\end{figure}

Fig.~\ref{fig:base_spmm} illustrates our basic design for implementing SpMM ($Y = A \times X$, where $A$ is sparse and $X$ is dense) using the low-level CSL language. The $N \times D$ dense matrix $X$ is distributed uniformly across a 2D grid of worker PEs. The $N \times N$ sparse matrix $A$ is represented in CSR format and is streamed into the system in blocks of \textit{max\_y\_chunk} rows at runtime. 
Specifically, the host sends batches of 32-bit integer column indices (\texttt{col\_idx}) and 32-bit floating point values (\texttt{value}) to the top-left PE. A single batch includes all of the column indices and values of \texttt{max\_y\_chunk} rows of $A$. \texttt{max\_y\_chunk} rows of output $Y$ are streamed back to the host when the results are available. While the illustration focuses on the case where $A$ is square, the design applies to non-square matrices as well.

In this design, there are three types of PEs each running their own code: \textit{routers}, \textit{workers}, and \textit{accumulators}. The PEs in the first column, called $routers$, are responsible for receiving matrix $A$ and \textit{dispatching non-zero elements} to the workers (or to other routers). Since tiles of $X$ are statically assigned to PEs, each non-zero element of $A$ is dispatched to the row of workers that contains the row of $X$ that must be multiplied by that element. For example, a non-zero
element $a_{mn}$ is dispatched to the row of workers storing row $X_{n*}$ and streamed to all the PEs in that row (to allow computing the dot-products across the whole row $X_{n*}$). As a result of this mapping, each row of PEs computes partial dot-products on a subset of the rows of $X$. Those partial dot-products are streamed south and progressively accumulated. An additional row of PEs, called \textit{accumulators}, receives the full dot-products (i.e., the computed \texttt{max\_y\_chunk} rows of $Y$) and streams them to the host. Below, we provide more details on the operation of the different types of PEs and the host code.

\begin{lstlisting}[language=csl,label=lst:spmm-router,caption=CSL pseudo-code for router PEs (basic design),float,frame=tb,numbers=left]
task recv_west(col_idx: u32) void {
    if(col_idx > last_col_idx){
        // end of row received
	    if (row_end == false){
	       @mov16(aout_dsd, DONE);
	       row_end = true;
        }
        if (col_idx == END_ROW) { row_end = false; }
        @mov32(aout_south_dsd, col_idx); // column index
        @fmovs(vout_south_dsd, vin_west_dsd); // value
    }else if (col_idx >= first_col_idx and col_idx <= last_col_idx){
   	  // send to worker PE
        @mov16(aout_dsd, @as(u16, col_idx - first_col_idx));
        
	    @fmovs(vout_dsd, vin_west_dsd); // value
    }
}

task recv_north(col_idx: u32) void {
    if(col_idx > last_col_idx){
        // end of row received
        if (row_end == false){
           @mov16(aout_dsd, DONE);row_end = true;
        }
        if (col_idx == END_ROW) {row_end = false;}
    }else if (col_idx >= first_col_idx and col_idx <= last_col_idx){
    	// send to worker PE
        @mov16(aout_dsd, @as(u16, col_idx - first_col_idx));
        @fmovs(vout_dsd, vin_north_dsd); // value
    }
}
\end{lstlisting}

\noindent $\blacksquare$ \textbf{Routers}: Router PEs serve as a buffer and pre-processor for incoming wavelets. Recall that one wavelet is a 32-bit value; thus, two wavelets compose a single nonzero value in $A$. One wavelet represents the column index and the other represents the value of the nonzero element. In our design, the row pointer does not need to be sent since the host attaches an \texttt{END\_ROW} character at the end of each stream of column indices and values to indicate the end of a row of $A$. The host additionally tracks which rows are sent to the device such that results are copied back to the appropriate region of the host output tensor. The router PEs execute the tasks shown in Listing~\ref{lst:spmm-router} (note that this code is simplified to showcase core functionality). Tasks \texttt{recv\_west} and \texttt{recv\_north} are triggered when a column index arrives and values are read in from either the host (through the \texttt{vin\_west\_dsd} fabric DSD) or from the north PE (through the \texttt{vin\_north\_dsd} fabric DSD), respectively. The top-left router is the only PE that will run the \texttt{recv\_west} task since it is the only PE that receives wavelets directly from the host.

\texttt{recv\_west} will either forward incoming column indices and values to the top row of worker PEs or will forward indices and values south to all other routers (arrow 1 in Fig.~\ref{fig:base_spmm}). \texttt{recv\_north} receives column indices and values from the router to the north of each router and determines if the index/value pair should be forwarded to the row of workers. Each row of workers is associated with a range of column indices to operate on which includes values within the interval $[\texttt{first\_col\_idx},\texttt{last\_col\_idx}]$. Thus, the $i$-th router at coordinate $(i, 0)$ in the PE grid forwards column indices in the range $[\texttt{i*max\_v\_per\_pe, (i+1)*max\_v\_per\_pe-1}]$ to the $i$-th row of workers (arrow 2 in Fig.~\ref{fig:base_spmm}). \texttt{max\_v\_per\_pe} is a compile-time parameter that determines how many rows of PEs are required for the SpMM kernel. Note that the column index forwarded to workers is an index relative to the first column index in the range represented as a 16-bit integer.

\begin{lstlisting}[language=csl, label=lst:spmm-worker,caption=CSL pseudo-code for worker PEs,float,frame=tb,numbers=left]
task recv_a(index: u16) void {
    // accumulate non-zero values unless at the end of the row    
    if (index != DONE){
        @fmacs(y_acc_dsd, y_acc_dsd, v_in_dsd, x[index]);
    }else{
        //if at the end of the row, moves to next row
        hsync_count = (hsync_count + 1) & (max_y_chunk - 1);
        if (hsync_count != 0){
            y_acc_dsd = @increment_dsd_offset(y_acc_dsd, 1, u32);
        }else{
            // if the chunk has been fully processed, reduce
            if (pidy == 0) {
                @fmovs(y_out_dsd, full_y_dsd, .{.async=true});
            }else{ 
                // listen from above and reduce down
                @fadds(y_out_dsd, y_in_dsd, full_y_dsd,.{.async=true});
            }
            // Reset the local buffer
            y_acc_dsd = @set_dsd_base_addr(y_acc_dsd, &y);
        }
    }
}
\end{lstlisting}

\noindent $\blacksquare$ \textbf{Workers}: Worker PEs perform multiplications and accumulations to generate partial results of $Y$. As stated above, the $i$-th row of workers only performs work on nonzero elements with column index in the range $[\texttt{i*max\_v\_per\_pe, (i+1)*max\_v\_per\_pe-1}]$. Prior to $A$ being streamed onto the CS-3, a local buffer \texttt{x} of length \texttt{max\_v\_per\_pe} stores a slice of a column of $X$. For simplicity, the pseudocode assumes that each column of PEs handles a column of $X$ (even though assigning multiple columns of $X$ to each column of PEs is possible). There are $D$ columns of worker PEs; thus, a row of workers operates on the same set of $A$ column indices, and a single worker stores a different column of $X$. Each worker computes \texttt{max\_y\_chunk} output elements of $Y$, though these are partial sums.

Listing~\ref{lst:spmm-worker} shows the CSL task code for each worker. Note that when a router sends an index/value pair to a row of workers, this pair is automatically forwarded to all workers in the row (arrow 2 in Fig.~\ref{fig:base_spmm}). \texttt{recv\_a} performs a floating-point multiply-accumulate on Line 4 for all nonzero values in the column index range for a row of $A$. When an \texttt{END\_ROW} index is received, the worker increments the output buffer pointer (Line 9) since the current $A$ row partial sum has completed. When \texttt{max\_y\_chunk} rows of $A$ have been processed, Lines 12-17 in Listing~\ref{lst:spmm-worker} perform an accumulation of \texttt{max\_y\_chunk} partial outputs of $Y$. The top row of workers forward their local $Y$ buffers south while all other rows of workers first receive the buffer from the row above, then add the incoming buffer to the local buffer, and lastly forward the result south. Recall that a single column of workers generates results for one column of $Y$. This north-to-south accumulation corresponds to arrow 3 in Fig.~\ref{fig:base_spmm}.

\noindent $\blacksquare$ \textbf{Accumulators}: Accumulator PEs receive \texttt{max\_y\_chunk} rows of $Y$ from the last row of worker PEs. Accumulators then stream \texttt{max\_y\_chunk} rows of $Y$ to the host (black arrows at the bottom of Fig.~\ref{fig:base_spmm}). Each accumulator streaming a set of columns of $Y$. Internally, accumulator PEs use a CS-3 FIFO buffer to asynchronously receive/send $Y$ elements.

\begin{lstlisting}[label=lst:spmm-host,caption=Pseudo-code for host (basic design),float,frame=tb,numbers=left]
def spmm(A):
	y_buff = tensor(max_rows,D)
	for c in max_rows/max_y_chunk:
		copy_h2d(src=A[c*max_y_chunk:(c+1)*max_y_chunk,:],
			dst=routers[0], stream=True, nonblock=True)
		copy_d2h(src=accumulators,
			dst= y_buff[c*max_y_chunk:
                        (c+1)*max_y_chunk,:],
			stream=True, nonblock=True)
\end{lstlisting}

\noindent $\blacksquare$ \textbf{Host code}: The host code is written in Python and pseudocode is shown in Listing~\ref{lst:spmm-host}. In the pseudocode, \texttt{max\_rows} is the largest number of rows of $X$ that can be accommodated on the CS-3 along the y-axis, and with $\texttt{max\_v\_per\_pe}=64$, $\texttt{max\_rows} = min(N,65536)$. There are $\texttt{max\_rows}/\texttt{max\_v\_per\_pe}$ rows of workers, thus for $\texttt{max\_v\_per\_pe}=64, \texttt{max\_rows}=65536$, 1024 rows of workers are used. Prior to calling \texttt{spmm(A)}, the host directly copies $X$ into the worker PEs, with each worker holding \texttt{max\_v\_per\_pe} $X$ elements. \texttt{spmm(A)} asynchronously streams \texttt{max\_y\_chunk} rows of $A$ at a time to the top-left router PE and assumes that $A$ is stored in CSR format. Not shown here is the addition of the \texttt{END\_ROW} character directly into the CSR in both the column index and value arrays. Listing~\ref{lst:spmm-host} additionally does not show the copying of the column index and value arrays separately. In the actual host code, column index and value pairs are interleaved such that they can be streamed together. After issuing the host-to-device streaming memory copy, the host issues a device-to-host streaming memory copy to receive \texttt{max\_y\_chunk} rows of $Y$.

\textbf{Design Considerations:} This SpMM design employs the dataflow nature of the CS-3 to process different columns and rows of $A$ in both pipeline-parallel and data-parallel fashion. The router pre-processing and filtering of incoming nonzero elements enables the matrix multiplication to perform computation only on nonzero values of $A$, avoiding unnecessary communication and computation present in a dense-dense matrix multiplication.
There are certain limitations of the CS-3 that drive our design choices. For instance, our design computes \texttt{max\_y\_chunk} rows of output at a time, introducing some synchronization in the SpMM kernel. This is due to the on-chip memory capacity of a PE. A single worker PE (and accumulator) must buffer \texttt{max\_y\_chunk} elements, thus for larger matrices, the CS-3 is limited to the maximum \texttt{max\_y\_chunk} that can be accommodated in the 46 KB of local memory. Additionally, the tile of $X$ stored in a single worker must be accommodated in local memory, thus \texttt{max\_y\_chunk} and \texttt{max\_v\_per\_pe} should be set to fit all necessary values in local memory.

\vspace{2mm}
\subsubsection{Optimized SpMM Design} 
\label{sec:spmm:optimized}
We introduce two optimizations to the basic design described above aimed to improve host-device communication and reduce synchronization costs.

\textbf{Optimization 1: Improving host-device communication} -- 
The initial SpMM design suffers from sub-optimal host-device communication bandwidth. With this design, column indices and values must be sent to only one router PE since worker PEs expect to perform multiply-accumulate operations for one output element at a time. Additionally, each row of workers only holds a portion of $X$ and must know when an incoming set of index/value pairs within the index range from a row of $A$ has completed, forcing all elements to be routed through the top-left router. The CS-3 itself is capable of extreme compute throughput, but the wafer must be fed data quickly enough to yield performance benefits over CPU and GPU. With the initial SpMM design, only one I/O channel on the west edge can be used to stream data onto the wafer, severely impeding performance.

\begin{figure}[b]
\vspace{-1mm}
\centerline{\includegraphics[width=0.9\linewidth]{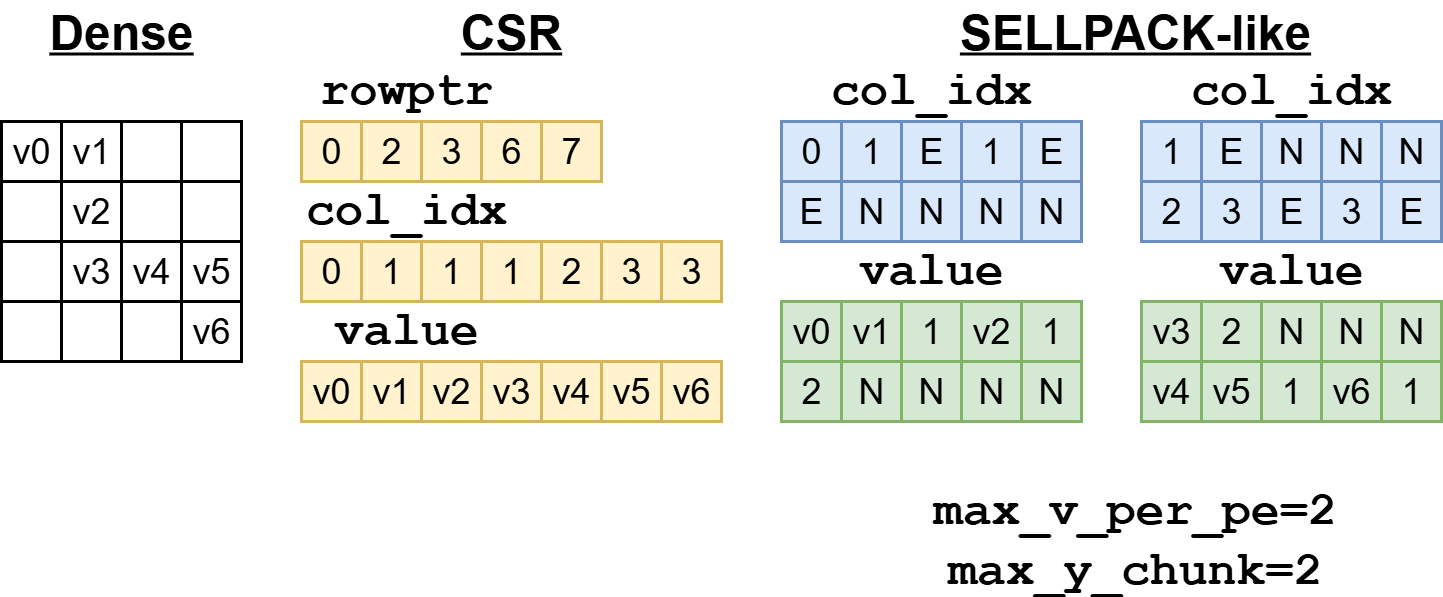}}
\caption{Storage formats for $A$. Column indices are shown in \texttt{col\_idx} and values are in \texttt{value}. ``E'' is the \texttt{END\_ROW} character and ``N'' is a null character.}
\label{fig:storage_formats}
\end{figure}

To improve the host-device communication bandwidth, we must be able to use more I/O channels. This requires index/value pairs to be sent directly to the router that forwards pairs to worker rows. If we are able to format the sparse matrix A such that nonzero column index/value pairs are sent directly to the router that consumes the pair, more I/O channels can be used.

Fig.~\ref{fig:storage_formats} shows three different formats for an example matrix $A$. From left to right, Fig.~\ref{fig:storage_formats} illustrates the dense format, the CSR format, and a Sliced Ellpack-inspired (``SELLPACK-like'') format. Sliced Ellpack (SELL) \cite{cusparse} is a state-of-the-art storage format for sparse matrix kernels on GPU. SELL slices $A$ into chunks and within each chunk, all nonzero values are left-aligned and right zero-padded. Thus, within a chunk, each row has as many elements as the maximum number of nonzeros of a row in that chunk. Each chunk is composed of both an index matrix and a value matrix. SELL generally requires more space than CSR, but data is organized in a more memory access-friendly way. On GPU, this format can improve memory coalescing.

The SELLPACK-like format we propose is structured with the SpMM kernel in mind. There are \texttt{}$N/\texttt{max\_y\_chunk}$ chunks. The $i$-th row of the SELLPACK-like format contains all of the nonzeros to be sent to the $i$-th row of workers and is populated based on the \texttt{max\_v\_per\_pe} index range as described in the initial SpMM design. Each row contains \texttt{END\_ROW} characters (\texttt{E} in Fig.~\ref{fig:storage_formats}) and is padded with \texttt{NULL} values (\texttt{N}). The SELLPACK-like format stores the run-length of \texttt{END\_ROW} characters in the \texttt{value} array, thus if a PE is receiving many consecutive empty rows, only a single \texttt{END\_ROW} index/value pair is needed. Within a chunk, there are as many rows of column index/value elements as there are PEs in the worker grid. Note that two arrays are required for each chunk: one for column indices and one for values. In the host code, these arrays are interleaved such that each router processes one pair of index/value at a time.

\begin{figure}[b]
\vspace{-1mm}
\centerline{\includegraphics[width=0.9\linewidth]{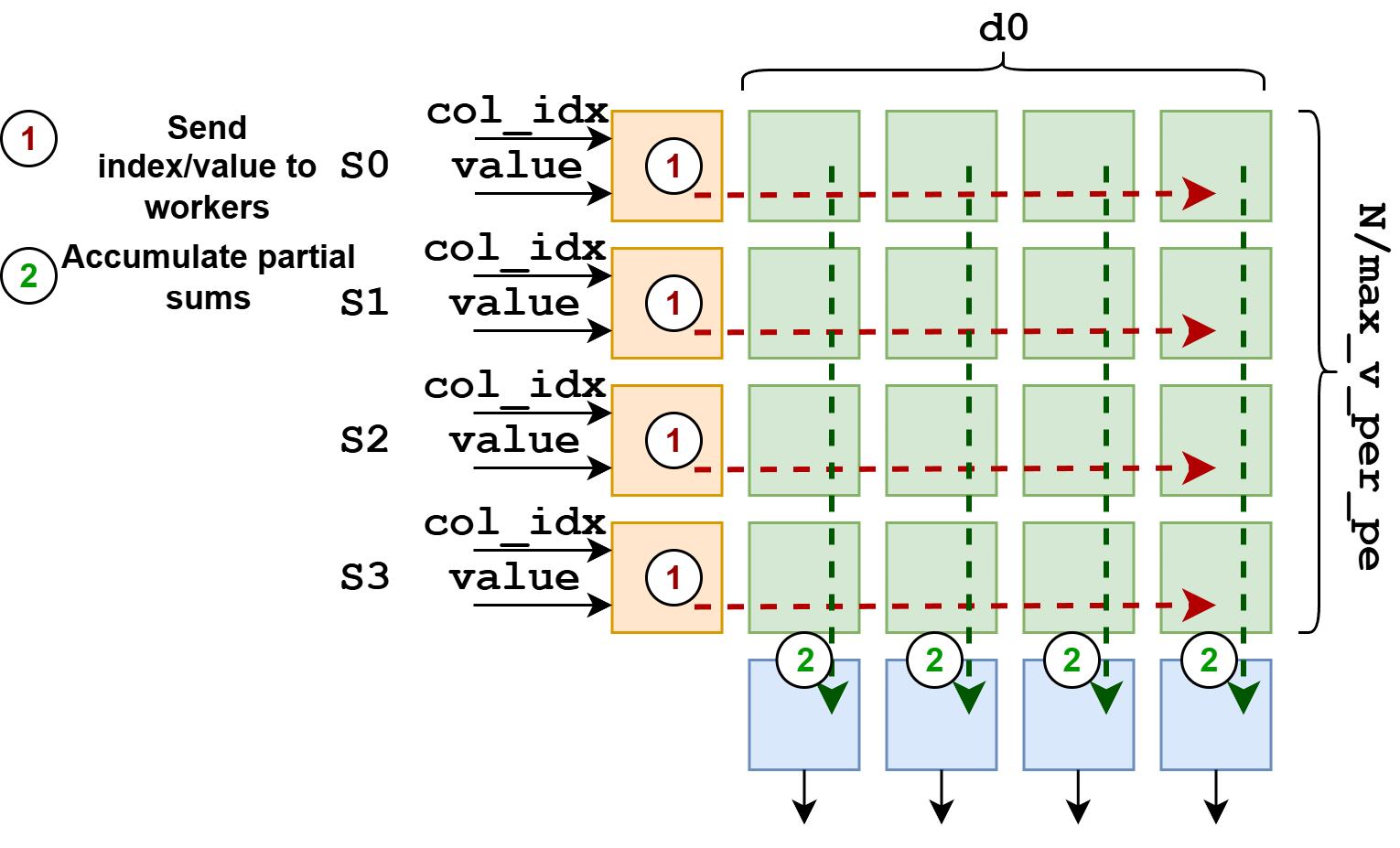}}
\caption{SpMM for CS-3 using SELLPACK-like format for $A$ ($D = d_0$)}
\label{fig:nnz_pad_spmm}
\end{figure}

Fig.~\ref{fig:nnz_pad_spmm} diagrams the SpMM kernel leveraging the SELLPACK-like format. Each stream \texttt{Sx} corresponds to the \texttt{x}-th row of a chunk of the SELLPACK-like $A$. Recall that one chunk corresponds to \texttt{max\_y\_chunk} rows of $A$. Each stream goes directly to the appropriate router and subsequently to the appropriate worker row. The updated router program is shown in Listing~\ref{lst:nnzpad-router}.

Compared to the initial SpMM design, the design in Fig.~\ref{fig:nnz_pad_spmm} eliminates the need for north-south communication from the top-left router to other routers. This can reduce back-pressure on the PE grid as well as improve host-device communication bandwidth. After the host has copied the SELLPACK-like chunk to the device, an asynchronous streaming memory copy from the accumulators back to the host is issued to collect \texttt{max\_y\_chunk} rows of $Y$.

\begin{lstlisting}[language=csl,label=lst:nnzpad-router,float, caption=CSL pseudo-code for router PEs (optimization 1),frame=tb,numbers=left]
task recv_west(col_idx: u32) void {
    if(col_idx!=EMPTY){
        if(col_idx==END_ROW){
            @mov16(aout_dsd, DONE);
            @fmovs(discard_buffer_dsd,vin_west_dsd);	
        }else{
            @mov16(aout_dsd, @as(u16, col_idx - first_col_idx));
            @fmovs(vout_dsd, vin_west_dsd);
        }
    }
}
\end{lstlisting}

\textbf{Optimization 2: Reducing Serialization} --
Recall the host code in Listing~\ref{lst:spmm-host}. For every chunk of the input matrix $A$, a streaming memory copy from device-to-host is issued to collect \texttt{max\_y\_chunk} rows of $Y$ back on the host. The CS-3 SDKRuntime library, which is used to run CSL programs on the host, serializes memory copy calls, thus the host-to-device memory copy on line 4 cannot proceed until the previous loop iteration device-to-host memory copy on line 6 completes. This leads to a partial serialization of the SpMM: the next chunk computation cannot proceed until the output of the previous chunk is on the host.

\begin{figure}[b]
\centerline{\includegraphics[width=0.65\linewidth]{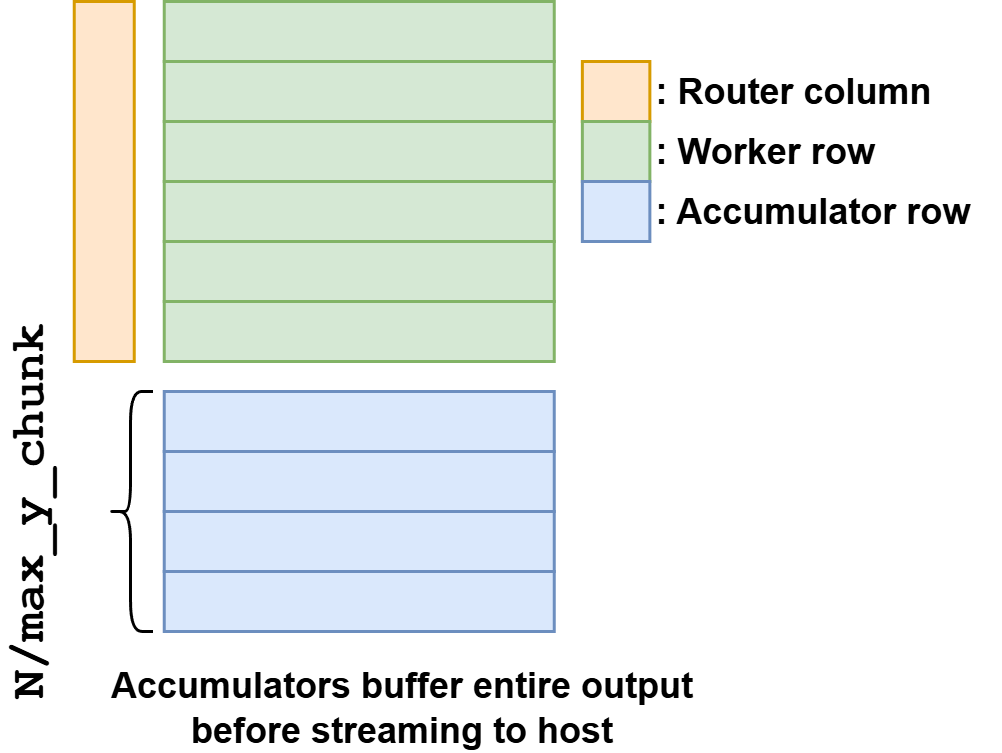}}
\caption{Buffering $Y$ on-chip with multiple accumulators}
\label{fig:multiple_acc}
\end{figure}

We extend the SpMM design to use multiple rows of accumulators, as shown in Fig.~\ref{fig:multiple_acc}. The modified host code is shown in Listing~\ref{lst:multipleacc-host}. Each row of accumulators can store $\texttt{max\_y\_chunk} \times D$ elements and the accumulator rows are filled with the entire output matrix $Y$. Thus, there are $N/\texttt{max\_y\_chunk}$ accumulator rows, or $\frac{\texttt{max\_rows}}{\texttt{max\_y\_chunk}}$ if $N$ is too large to fit on the CS-3. On the host, only one device-to-host streaming memory copy is needed and is called after all host-to-device memory copies are issued. As such, only the host-to-device memory copies are serialized, enabling better compute overlap for PEs. One additional benefit is the use of more I/O channels to stream data to the host.

\begin{lstlisting}[label=lst:multipleacc-host,caption=Pseudo-code for host (optimization 2),float,frame=tb,numbers=left]
def spmm(A):
    y_buff = tensor(max_rows,D)
    for c in max_rows/max_y_chunk:
        copy_h2d(src=A[c*max_y_chunk:(c+1)*max_y_chunk,:],
                dst=routers, stream=True, nonblock=True)
    copy_d2h(src=accumulators, dst= y_buff[:,:], 
            stream=True, nonblock=False)
\end{lstlisting}

\subsection{Sampled Dense-Dense Matrix Multiplication on the CS-3}

\begin{figure}[b]
\vspace{-2mm}
\centerline{\includegraphics[width=0.75\linewidth]{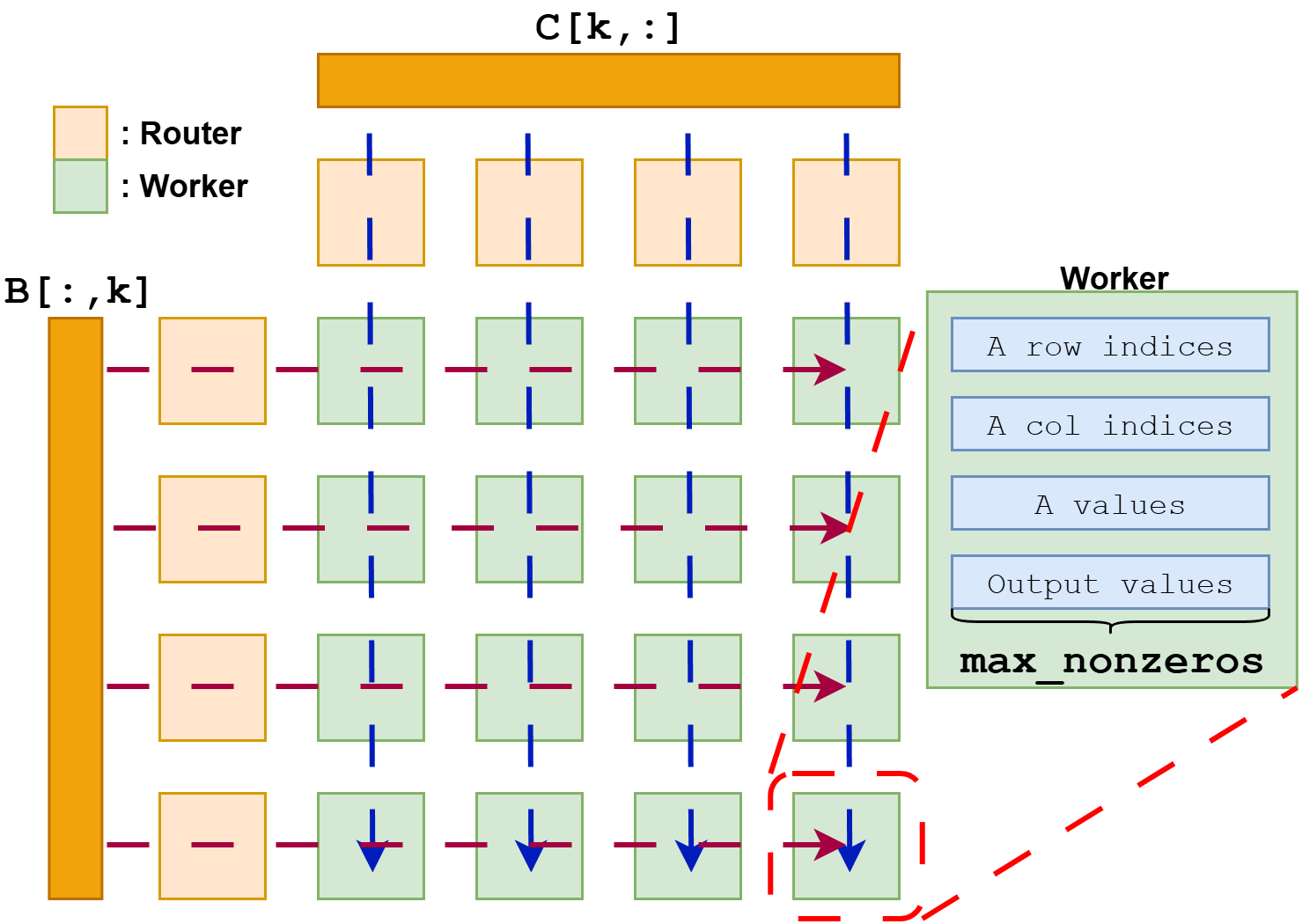}}
\caption{Sampled Dense-Dense Matrix Multiplication (SDDMM) kernel for CS-3 execution to compute $Y=A \odot(BC)$ where $A$ is sparse and $B,C$ are dense.}
\label{fig:sddmmm}
\end{figure}
Formally, SDDMM computes $Y=A \odot(BC)$ where $A$ is sparse and $B,C$ are dense. Often, calculating the dense product of $B$ and $C$ is costly, thus SDDMM focuses only on computing $Y$ according to the sparsity pattern of $A$. 

Fig.~\ref{fig:sddmmm} illustrates our proposed design for SDDMM on the CS-3. There are two types of PEs: \textit{routers} and \textit{workers}. Router PEs, located along either the north edge or west edge, forward elements of $B$ and $C$ from the host to the worker PEs. Worker PEs receive elements from either the north or west directions and compute the elements of $Y$ only where elements of $A$ are nonzero. Each worker holds the nonzeros of a tile of $A$, sized according to the parameters \texttt{local\_width} and \texttt{local\_height} (the number of columns and rows of $A$, respectively). The parameter \texttt{max\_nonzeros} denotes the number of elements of $A$ that can be held in a single worker. Since $A$ is sparse, \texttt{max\_nonzeros} can be set to be lower than \texttt{local\_width} $\times$ \texttt{local\_height}. $A$ is stored in a COO format, thus the row index, column index, and values are all in local memory as well as the output buffer for $Y$.

To compute the SDDMM, the host streams in a column of $B$ to the west column of routers and a row of $C$ to the north row of routers. Elements are propagated through all worker PEs according to the directions showed by the dashed arrows in Fig.~\ref{fig:sddmmm}. After the workers have finished processing the current column/row of $B$ and $C$, the host streams the next column/row. Once all elements of $B$ and $C$ have been sent, the output buffers storing $Y$ elements are streamed back to the host. Row index, column index, and value arrays on each worker are padded with NULL values if the nonzeros of a given tile of $A$ are less than \texttt{max\_nonzeros}.

\section{Evaluation}
\subsection{Methodology}
To evaluate the proposed CS-3 kernels, we compare SpMM and SDDMM CS-3 designs with CPU results. The baseline CPU is an AMD EPYC 9354P with 188 GB of memory. We use the PyTorch sparse library with PyTorch 2.10 and SciPy to implement graph operations on CPU. The CS-3 system uses CS SDK version 1.4 for CSL programming and runtime. Experiments report the timing for random sparse and dense matrices with $K=N$, and $N$ varying from 2048 to 65536. We set $D=256$ (a commonly used size in state-of-the-art GNN models \cite{ogb}). Each figure reports the results averaged across 50 runs.

\subsection{Adjacency Matrix Footprint}
\begin{figure}[t]
\centerline{\includegraphics[width=0.6\linewidth]{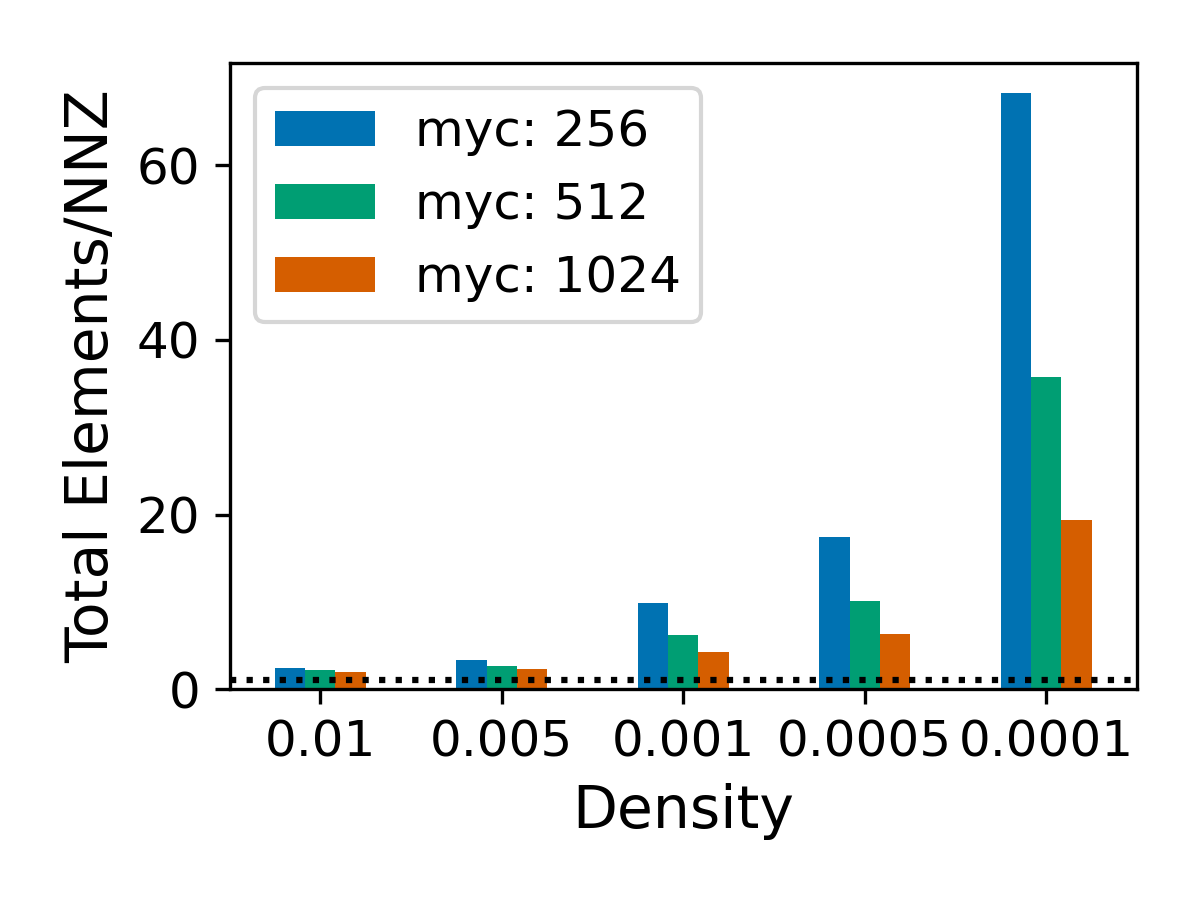}}
\vspace{-2mm}
\caption{Ratio of the total number of elements sent from the host using the SELLPACK-like format to the number of nonzeros for varying density values averaged across random $N \times N$ matrices, for $N=16384, 32768, 65536, 131072, 262144$.}
\label{fig:results_adj_footprint}
\end{figure}

Fig.~\ref{fig:results_adj_footprint} reports the ratio of the total number of elements sent by the host to the device with the SELLPACK-like format to the number of nonzero elements sent with a CSR format. Recall that, in our SpMM design, only column indices and values are sent to the CS-3 wafer. Each subplot corresponds to a different \texttt{max\_y\_chunk} setting as denoted on the left-hand side (\textit{myc}). Each series corresponds to a different number of rows of $A$.

As shown in Fig.~\ref{fig:results_adj_footprint}, the SELLPACK-like format has a larger memory footprint relative to CSR as the sparsity increases (density decreases). This is due to two factors: 1) the \texttt{END\_ROW} characters required to denote the end of a sequence of nonzeros for a row of $A$, and 2) the \texttt{NULL} padding to ensure every router PE receives the same number of elements streamed from the host. For extremely low density, such as $10^{-4}$ density (or 99.99\% sparsity), the SELLPACK-like format requires 20 to 80$\times$ more space than CSR. \texttt{max\_y\_chunk} also impacts the memory footprint: larger \texttt{max\_y\_chunk} reduces the memory footprint. When $N=262144, \textit{myc}=1024$ the SELLPACK-like format requires $\approx$1.1 GB and when $N=262144, \textit{myc}=256$, this format requires $\approx$4.1 GB. CSR for $N=262144$ requires $\approx$53 MB. For higher density, such as 0.01 density (99\% sparsity), SELLPACK-like storage converges to CSR, occupying only $\approx1.5\times$ the memory of CSR. Note that if $A$ were stored in a dense format and $N=262144$, the matrix would occupy 256 GB. Thus, the SELLPACK-like format is still significantly more space efficient than a dense format for sparse matrices.

\subsection{SpMM results}
\begin{figure}[h]
\centerline{\includegraphics[width=0.9\linewidth]{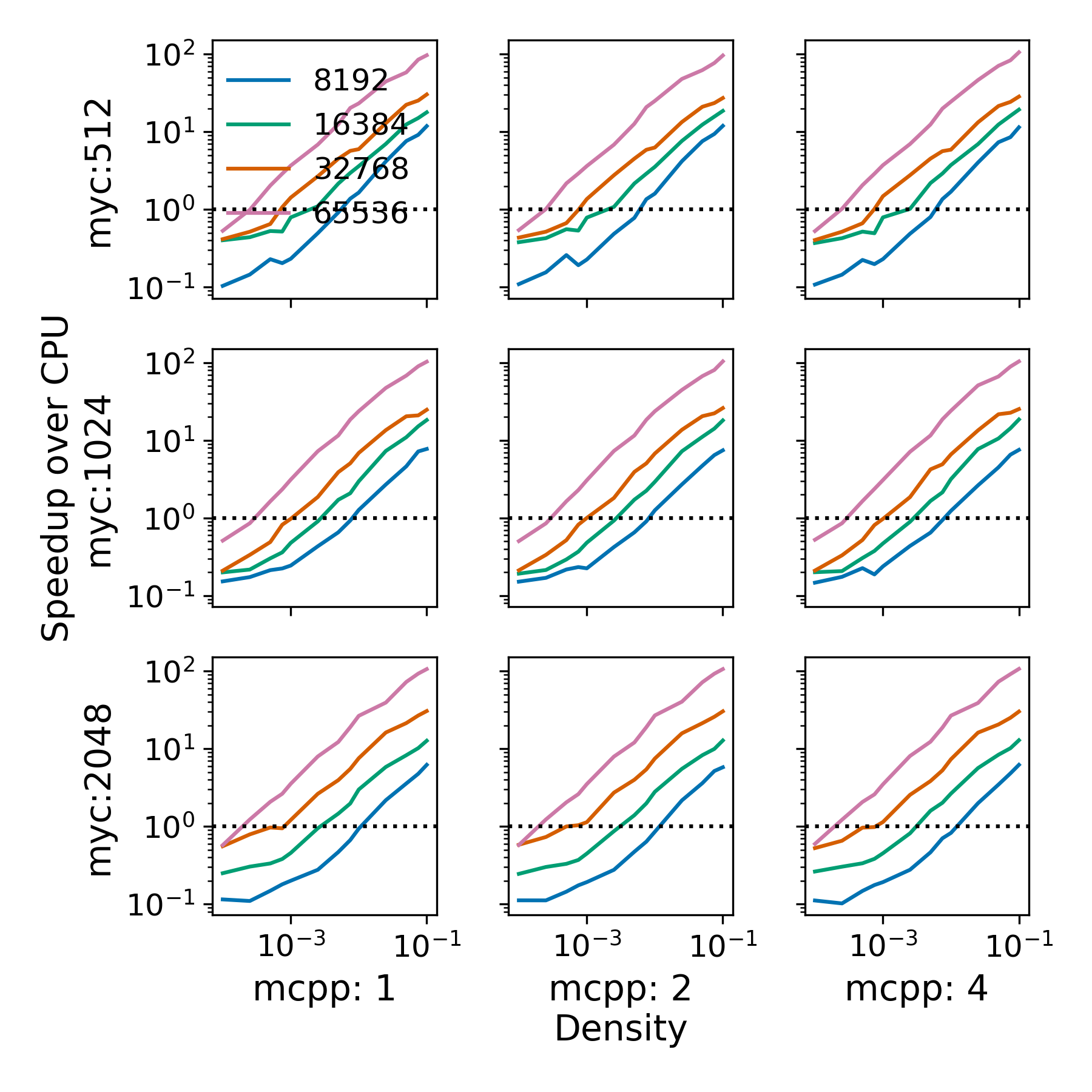}}
\caption{Speedup of CS-3 over CPU for the SpMM computation $Y=A \times X$, where $A \in\mathbb{R}^{N \times N}$ and $X \in\mathbb{R}^{N \times D}$. Each subplot corresponds to a different \texttt{max\_y\_chunk} (\textit{myc} varies along rows) and \texttt{max\_col\_per\_pe} (\textit{mcpp} varies along columns). Each series corresponds to a different $N$;  $D=256$.}
\vspace{-1mm}
\label{fig:results_spmm}
\end{figure}

Fig.~\ref{fig:results_spmm} reports the speedup of the CS-3 relative to CPU for the SpMM computation $Y=A \times X$, where $A \in\mathbb{R}^{N \times N}$ and $X \in\mathbb{R}^{N \times D}$. The density of $A$ is varied along the x-axis, and each series corresponds to a different $N$. Each row corresponds to a \texttt{max\_y\_chunk} (\textit{myc}), and each column corresponds to a \texttt{max\_col\_per\_pe} (\textit{mcpp}). We fix $D=256$ for $X$.

At a high level, the CS-3 can outperform the CPU by more than 100$\times$, though CS-3 performance is very sensitive to density. Higher-density $A$ yields significantly better performance than  CPU, while lower density (extreme sparsity) can result in worse performance than the CPU. This is due to a few factors: 1) lower density does not provide enough work to fully utilize the wafer PEs, 2) lower density has a higher relative overhead of streaming data to and from the device, and 3) the degree of pipeline parallelism is more limited with fewer nonzeros.

Performance is relatively consistent across varying \textit{mcpp}, with the speedups across \textit{mcpp} being in the range of $0.1$ to $100\times$. This finding is significant for mapping multiple SpMM kernels: higher \textit{mcpp} utilizes fewer PEs while maintaining performance comparable to lower \textit{mcpp}. 
When varying \textit{myc}, larger $N$ matrices benefit from larger \textit{myc}, with a 10--15\% improvement in execution time for \textit{myc}=2048 compared to \textit{myc}=512, while smaller \textit{myc} performs better for smaller $N$.

\subsection{SDDMM results}
\begin{figure}[t]
\centerline{\includegraphics[width=0.75\linewidth]{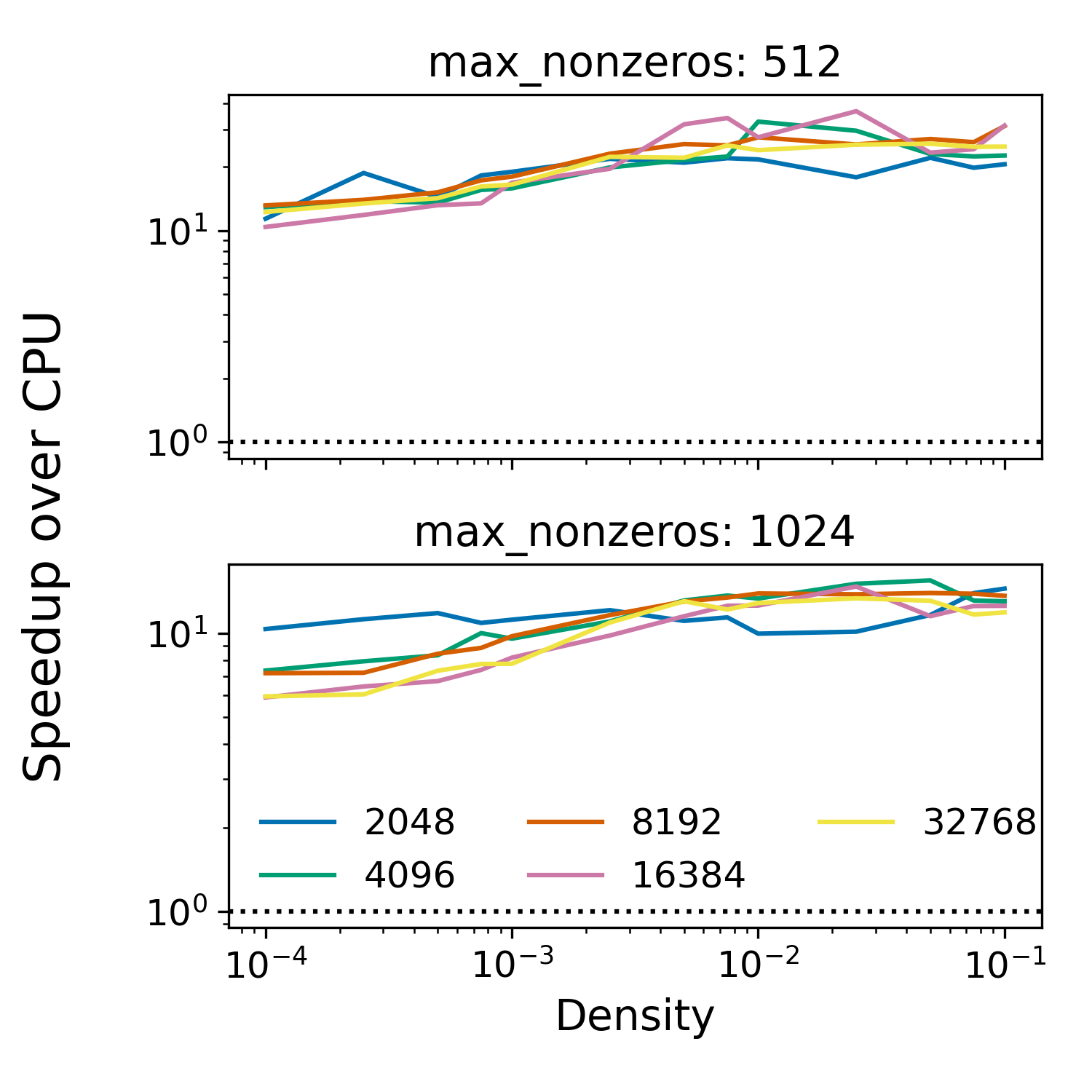}}
\caption{Speedup of the CS-3 over CPU for SDDMM. Each series plots a different $N$ while each subplot shows results for a different values of \texttt{max\_nonzeros} ($mnz$). Density is varied along the x-axis.}
\label{fig:results_sddmm}
\vspace{-2mm}
\end{figure}
Fig.~\ref{fig:results_sddmm} reports the speedup of CS-3 over CPU for the SDDMM computation $Y=A \odot (BC)$, with $A \in\mathbb{R}^{N \times N}$, $B \in\mathbb{R}^{N \times D}$, and $C \in\mathbb{R}^{D\times N}$. 
In these experiments, $D$ is set to 2 (a common setting for graph attention networks), and $N$ varies according to each series. The parameter \texttt{max\_nonzeros}, or \textit{mnz}, indicates the size of the nonzero buffer space for elements of $A$ on each worker PE. Recall that each worker PE holds the nonzeros for a tile of $A$; for this experiment, each worker holds the nonzeros for a $64 \times 64$ tile of $A$.

CS-3 performance is generally at least an order of magnitude faster than CPU performance, with performance slightly increasing as the density increases. We observe a shallower slope for performance improvement since the padding required for worker $A$ buffers increases device-to-host memory copy pressure, while density only influences the amount of computation performed. Since the CS-3 has high compute density and SDDMM is memory-bound, the performance bottleneck is memory copies that are dependent on $mnz$. We additionally observe that lower $mnz$ can yield better performance ($mnz=512$ can be up to $3\times$ faster than $mnz=1024$). Since the output buffer storing $Y$ is dependent on $mnz$, a larger $mnz$ indicates more data movement from the device to the host. For hyper-sparse matrices, a lower $mnz$ could further improve CS-3 performance relative to the CPU. 

\section{Conclusion and Future Work}
In this work, we proposed novel SpMM and SDDMM kernels for the CS-3 and explored SpMM optimizations to improve host-device communication. Our evaluation shows that our kernels can outperform the CPU by up to 100$\times$ for SpMM, while SDDMM can be up to 20$\times$ faster. However, the CS-3 is not efficient for high sparsity ($>99\%$), limiting the practical benefits of our kernels for large, hyper-sparse matrices. Future work includes: (1) investigating SpMM and SDDMM designs and mapping strategies better suited to hyper-sparse matrices, (2) exploring additional sparse matrix formats, and (3) reducing host-device communication overhead for back-to-back matrix operations by keeping the result matrix resident on the CS-3.


\bibliographystyle{IEEEtran}
\bibliography{FullBib, refs}

\end{document}